\documentclass[fleqn,usenatbib]{mnras}

\usepackage{newtxtext,newtxmath}
\usepackage{subcaption}

\usepackage[T1]{fontenc}

\DeclareRobustCommand{\VAN}[3]{#2}
\let\VANthebibliography\thebibliography
\def\thebibliography{\DeclareRobustCommand{\VAN}[3]{##3}\VANthebibliography}

\usepackage{graphicx}	
\usepackage{amsmath}	
\usepackage{bm}

\title[Maximum Energies of Electrons ]{Maximum possible energies of electrons accelerated in
magnetospheres of rotating black holes}

\author[N. Nikuradze and Z. N. Osmanov]{
N. Nikuradze,$^{1}$\thanks{E-mail: nniku21@freeuni.edu.ge (NN)}
Z. N. Osmanov,$^{1,2}$\thanks{E-mail: z.osmanov@freeuni.edu.ge (Z.O.)}
\\
$^{1}$School of Physics, Free University of Tbilisi, 0159, Tbilisi,Georgia\\
$^{2}$E. Kharadze Georgian National Astrophysical Observatory, Abastumani 0301, Georgia\\
}

\date{Accepted XXX. Received YYY; in original form ZZZ}

\pubyear{\the\year{}}

\begin{document}
\label{firstpage}
\pagerange{\pageref{firstpage}--\pageref{lastpage}}
\maketitle

\begin{abstract}
Our aim is to evaluate the maximum attainable energies of electrons accelerated by means of the magneto-centrifugal mechanism. We examine how the range of maximum possible energies, as well as the primary limiting factors, vary with black hole mass. Additionally, we analyse the dependence of the maximum relativistic factor on the initial distance from the black hole and its spin factor in the range 0.1 - 0.2. We model the acceleration of electrons on rotating magnetic field lines and apply several constraining mechanisms: the inverse Compton scattering, curvature radiation, and the breakdown of the bead-on-the-wire approximation. As a result, the maximum Lorentz factors for electron acceleration vary with the type of black hole. For stellar-mass black holes, electrons can be accelerated up to the Lorentz factors $1.3 \times 10^{3} - 1.3 \times10^{4}$ with only co-rotation constrain affecting the maximum relativistic factor; In intermediate-mass black holes, the Lorentz factors are in the interval $1.3 \times 10^4 - 1.1 \times 10^{5}$; For the supermassive black holes the Lorentz factors range from $7 \times 10^{4}$ to $1.7 \times 10^{5}$; while the ultra-massive black hole located at the center of Abell 1201 can accelerate electrons up to $10^{6}$ with both the co-rotation and Inverse Compton in Thomson regime determining the final Lorentz factor for the last three categories.
\end{abstract}




\section{Introduction}

Mainly, there are three types of black holes based on their mass: stellar-mass black holes with $M_{BH} < 10^2M_\odot$, intermediate-mass black holes (IMBH) with $10^2M_\odot < M_{BH} < 10^5M_\odot$, and supermassive black holes (SMBH) with $M_{BH} > 10^5M_\odot$. Sometimes, another group of ultramassive black holes is added, such as the one in the center of Abell 1201 with $M_{BH} = 3.27 \times 10^{10}M_\odot$, which was discovered via gravitational lensing \citep{Nightingale2023}. Although stellar-mass and supermassive black holes have been detected, the existence of intermediate-mass black holes (IMBHs) remains uncertain. There are potential locations where IMBHs could exist, such as one that could reside at the center of the 47 Tucanae globular cluster, as suggested by \cite{Kızıltan2017}. However, this has yet to be definitively confirmed. Proposing a new method of measuring the mass of a black hole would help us to unravel the mystery concerning IMBHs or to verify the old data. Our method links the mass of a black hole to its ability to accelerate electrons with its magnetic field.

In cosmic ray astrophysics, one of the major problems is understanding the mechanisms that provide high energies. The so-called Fermi acceleration \citep{Fermi1949} and its modifications \citep{Bell1978I, Bell1978II, Catanese1999} although might explain particle acceleration to relativistic energies, as it turned out, the mechanisms are efficient if the particles are already pre-accelerated \citep{Rieger2000}. The efficiency of the Blandford-Znajeck process \citep{Blandford1977}, is significantly reduced by means of the screening effects.

The prevailing characteristic of black holes is their tendency to rotate, accompanied by the co-rotation of their magnetic field lines. Electrons, caught within the magnetic field, undergo a trajectory governed by the Lorentz force, resulting in a spiral path. The synchrotron emission rapidly diminishes the electron's perpendicular momentum, confining its motion primarily along the magnetic field lines. In particular, one can straightforwardly estimate that the synchrotron cooling time-scale, $\tau_s\simeq\gamma m_ec^2/P_s$ ($\gamma\approx10^{3}$ and $m_e$ are the electron's relativistic factor and mass respectively, $c$ denotes the speed of light,  $P_s\simeq 2e^4B^2\gamma^2/(3m_e^2c^3)$ is the synchrotron emission power, $e$ is the electron's charge and $B\approx2.35\times10^3G$ denotes the magnetic induction and is calculated in the section 3.1 using Bondi accretion model \citep{Bondi1952}) is of the order of $10^{-2}s$, which is smaller than the rotation period of a black hole, $P \simeq 2\pi (GM_{BH})/c^3\simeq 1230s$ where $M_{BH} = 10^{6}M_{\odot}$. This means that the particles lose their perpendicular momentum soon after they accelerate, transit to the ground Landau state, and proceed to slide along the field lines. The rotation of the magnetic field induces a centrifugal force on the electrons, leading to their acceleration. However, this acceleration is limited due to the counteracting effects of mechanisms such as inverse Compton scattering and curvature radiation, which cause the energy loss process. Consequently, electrons reach a maximum energy level determined by the equilibrium between acceleration and energy loss mechanisms. Furthermore, electrons may attain a critical energy threshold at which they break off the magnetic field line, setting an additional constraint on their maximum energy. A similar phenomenon has been studied in active galactic nuclei (AGN) by \cite{Osmanov2007}. The authors investigated the efficiency of centrifugal acceleration of electrons as a possible mechanism for the generation of ultra-high energy $\gamma$ -ray emission in TeV blazars and studied the phenomenon for various inclination angles of magnetic field lines. Additionally, it has been shown that electrons can attain $\gamma_{max}\approx10^8$ with the constraining factor of inverse Compton scattering.  Moreover, the acceleration process has been studied by \cite{Osmanov2009} and reviewed by \cite{Osmanov2021} for high-energy $\gamma$-ray pulsars. This study focused on three constraining factors: inverse Compton scattering, curvature radiation, and co-rotation. It has been shown that the maximum Lorentz factor of $10^7$ can be achieved. While the effectiveness of centrifugal acceleration as a mechanism in black holes has been previously explored, studies have not been focused on black hole mass and the initial radial coordinate as primary parameters. Our research addresses this gap by examining the acceleration mechanism across a broad spectrum of black hole masses. Particularly, we focus on intermediate-mass black holes (IMBHs) and the ultra-massive black hole (UMBH) at the centre of Abell 1201. We assess the efficiency of this acceleration process within black hole magnetospheres with masses spanning $1-10^{9}M_{\odot}$ and for the black hole at the centre of Abell 1201 with mass $3.27\times10^{10}M_{\odot}$. In the framework of the manuscript, we examine the so-called open field lines having very large values of curvature radii. It is clear that this is a narrow class of field lines, which we assume to be located in the equatorial plane. But not all field lines located in the mentioned zone are straight. On the other hand, even these straight field lines nearby the light cylinder (LC) area (a hypothetical surface, where the linear velocity of rigid rotation exactly equals the speed of light) will inevitably twist, lagging behind the rotation \citep{Osmanov2008} when the process of particle acceleration is terminated very rapidly. Therefore, for studying the dynamics of acceleration, we examine rectilinear magnetic field lines almost up to the LC zone, and only in the very vicinity of the mentioned area, the magnetic field lines are swept back. 

Several factors might influence the acceleration process. In particular, significant constraints might occur from the inverse Compton scattering, curvature radiation, and co-rotation effects. Additionally, we analyse how the relativistic factor and the main constraining factor depend on the distance from the BH.

\section{Acceleration process}
   In this section, we consider a black hole with mass $M_{BH}$ that rotates at angular velocity 
   \begin{equation}
       \Omega = \frac{c^3}{GM_{BH}} \frac{a}{2(1+\sqrt{1-a ^ 2} )}
       \label{RotVel}
   \end{equation} 
   where $ a = Jc/GM_{BH}^2$ is a spin parameter of a black hole \citep{Thorne1986,Shapiro1983}. We consider electrons moving along the magnetic field line in the rotating plane. The motion of electrons along curved trajectories was thoroughly investigated by \cite{Rogava2003}. Although our study focuses on straight magnetic field lines in the horizontal plane, the work by \cite{Rogava2003} serves as a crucial foundational reference for our research. Electron's motion is described in polar coordinates where coordinates, velocity, and acceleration can be written as:
   \begin{equation}
       \Vec{r} = r \cdot \Vec{e_r}
   \end{equation}
   \begin{equation}
       \Vec{\upsilon} =\frac{d\Vec{r}}{dt} =  \dot{r} \cdot \Vec{e_r} + r\dot{\theta} \cdot \Vec{e_{\phi}}
   \end{equation}
   \begin{equation}
       \Vec{a} =\frac{d\Vec{\upsilon}}{dt} = \left(\Ddot{r} - \Omega^2r\right ) \cdot \Vec{e_r} +2\dot{r}\dot{\theta} \cdot \Vec{e_\phi} 
   \end{equation}
\begin{figure}
    \centering
    \includegraphics[width=0.35\textwidth]{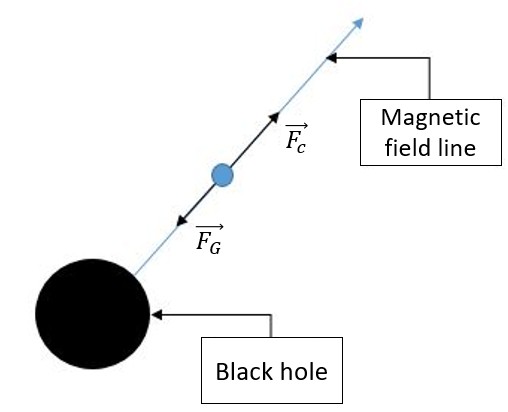} 
    \caption{Force diagram in the rotating frame}
    \label{fig:phenomenon}
\end{figure}
   \begin{equation}
       \gamma =\frac{1}{\sqrt{1-\frac{\dot{r}^2+r^2\dot{\theta}^2}{c^2}}}
   \end{equation}
   For the force, one has the following expression
\begin{equation}
\begin{split}
    \Vec{F} &= \frac{d(\gamma m_e \Vec{\upsilon})}{dt}
    &= m_e \left( 
        \frac{\dot{r}^2 (\ddot{r} + \Omega^2 r)}{c^2 \left(1 - \frac{\upsilon^2}{c^2} \right)^{\frac{3}{2}}} 
        + \frac{\ddot{r} - \Omega^2 r}{\left(1 - \frac{\upsilon^2}{c^2} \right)^{\frac{1}{2}}} 
    \right) \Vec{e_r} \\
    &\quad + F_{\theta} \Vec{e_\theta}
\end{split}
\end{equation}

    The only force acting on the electron in the radial direction is gravitational (Fig. \ref{fig:phenomenon}). In the case of a small spinning factor, most of the acceleration process happens far from the BH, hence, the gravitational force is negligible. Therefore, $F_r = 0$ (in a lab frame) gives rise to the final equation describing the acceleration process of the electron.
    \begin{equation}
    \label{acc}
        \ddot{r} = \Omega^2r\frac{1 - \frac{\Omega^2r^2}{c^2} - \frac{2\dot{r}^2}{c^2}}{1 - \frac{\Omega^2r^2}{c^2}}
    \end{equation}
\begin{figure}
     \centering
     \begin{subfigure}{0.35\textwidth}
         \centering
         \includegraphics[width=\textwidth]{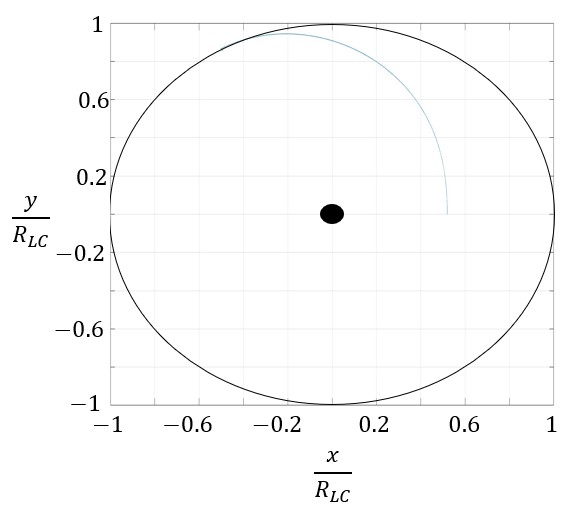}
         \caption{}
         \label{fig:trajectory}
     \end{subfigure}
     \hfill
     \begin{subfigure}{0.35\textwidth}
         \centering
         \includegraphics[width=\textwidth]{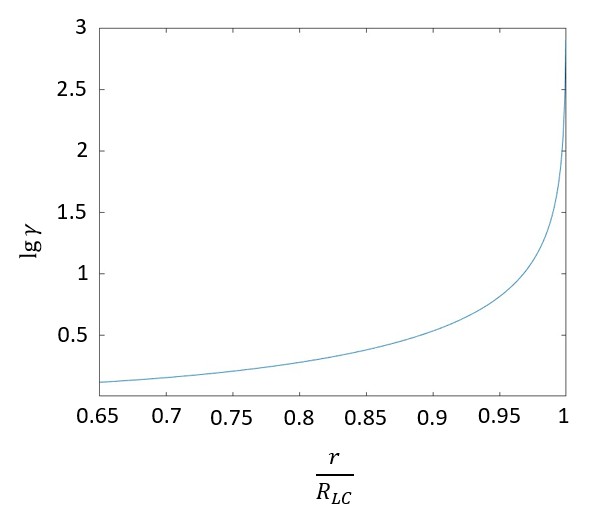}
         \caption{}
         \label{fig:RFonDist}
     \end{subfigure}
     \hfill
     
        \caption{(a) Electron trajectory (Blue curve), Black disk in the middle is the BH and outer black circle is the LC (b) The Lorentz factor dependence on the radial distance}
        \label{fig:firstres}
\end{figure}
     Generally speaking, the aforementioned equation implies that the particle co-rotates rigidly up to the LC area, which is not correct on LC, because no massive particle can reach the speed of light. In real scenarios, several important factors will inevitably terminate the co-rotation process.
    
    In particular, after solving Eq. (\ref{acc}), one can see from Fig. \ref{fig:trajectory} that the particle reaches the LC zone asymptotically, and this behaviour takes place until the energy gained from the centrifugal force overwhelms magnetic energy and consequently the electron breaks out from the field line. The co-rotation condition is discussed in detail in section 3.1. Tangential velocity of the electron is less than the speed of light so $\Omega r < c \Rightarrow r < \frac{c}{\Omega} \Rightarrow r_{max} = \frac{c}{\Omega}\equiv R_{LC}$ When the electron approaches the LC, its tangential velocity $\upsilon_{\theta} \rightarrow c$. Therefore, the radial velocity approaches zero. So, if the electron is bound to the straight magnetic field line, it can never leave the LC area. From Fig. \ref{fig:RFonDist}, we can see that the Lorentz factor of the electrons increases rapidly near the LC, and it would have gone to infinity if no constraining factors occurred. A very interesting analytical result can be inferred from the equation of motion. It appears that the dependence of the Lorentz factor on radial distance is as follows:
    \begin{equation}
        \label{LFonDistance}
        \gamma = \gamma_0\frac{1-\frac{r_0^2}{R_{LC}^2}}{1-\frac{r^2}{R_{LC}^2}}
    \end{equation} 
    where $\gamma_0$ and $r_0$ are initial Lorentz factor and radial distance respectively \citep{Rieger2011,Osmanov2016}. In the manuscript we focus on a small spinning parameter ($a =( 0.1-0.2)$ interval is used in the results sections), so that $R_{LC} >> R_{s}$ and since the most of the acceleration process happens near the LC, gravitational force becomes negligible, therefore Eq. (\ref{LFonDistance}) becomes valid to describe the dynamics of the accelerating particle.

\section{Constraining factors}
    In this section, we examine various types of limiting factors that restrict the maximum attainable energies of particles.
    \subsection{Co-Rotation Constraint}
    In the acceleration process, we have considered that the electron follows the magnetic field line. However, if the energy of the electron becomes too large, and as we have discussed in the introduction, the field lines are swept back, resulting in the termination of the acceleration process. Therefore, the electron reaches its maximum relativistic factor. Co-rotation constraint is satisfied if the magnetic field energy density, $B^2/8\pi$, exceeds the plasma energy density $\gamma n_{GJ}m_ec^2$, where $B$ denotes the magnetic field induction $B = 2\sqrt{\pi\rho\upsilon^2}$, where $\rho$ and $\upsilon$ are calculated through the Bondi spherical accretion model \citep{Bondi1952} resulting in the following expression of the magnetic field:
    \begin{equation}
        \label{Eq: MagneticField}
        B\approx\left(\frac{\pi\rho_{\infty}}{\sqrt{2}u_{\infty}^3}\right)^{\frac{1}{2}}\left(\frac{GM_{BH}}{r}\right)^{\frac{5}{4}}
    \end{equation} 
where we have used values  $\rho_{\infty}\approx10^{-24}g/cm^3$ and $u_{\infty}\approx10^6cm/s$  being the hydrogen density and sound speed very far from the BH, respectively, which are typical for the interstellar medium 
\citep{Shapiro1983}.  

\begin{equation}
    n_{GJ} = \frac{\Omega B}{2\pi e c \left(1 - \frac{r^{2}}{R^{2}_{LC}} \right)}
\end{equation}

$n_{GJ}$ denotes the Goldreich–Julian number density of electrons \citep{goldreich1969}, and the magnetic induction on the LC surface is of the order of $2.35 \cdot 10^3\, \mathrm{G}$.
From Eq.\ref{LFonDistance} we can substitute $\left(1 - \frac{r^{2}}{R^{2}_{LC}}\right)^{-1} = \gamma \gamma_{0}$,
Therefore, the expression for the Goldreich–Julian number becomes:

\begin{equation}
    n_{GJ} = \gamma \gamma_{0} \frac{\Omega B}{2\pi e c}
\end{equation}
    For the co-rotation constraint, one can write
    \begin{equation}
        \frac{B^2}{8\pi}\geq\gamma n_{GJ}m_ec^2 \Rightarrow \gamma \leq \sqrt{\frac{Be}{4 \gamma_{0} \Omega m_ec} } \label{Eq:RFmax}
    \end{equation}
    Since the electron must rotate with the magnetic field line, it has at least tangential velocity, which makes the Lorentz factor greater than its minimum value.
    \begin{equation}
        \gamma = \left( 1-\frac{\dot{r}^2}{c^2}-\frac{r^2\Omega^2}{c^2}\right)^{\frac{1}{2}} \geq \left(1-\frac{r^2\Omega^2}{c^2}\right)^{\frac{1}{2}}
    \end{equation}
    \subsection{Inverse Compton effect}
    The area inside the LC of the BH is filled with thermal photons radiated from an accretion disk. In due course of motion the electrons collide with photons and lose energy (Inverse Compton (IC) effect). In general, the IC scattering might occur in two extreme regimes, in the so-called Thomson regime for very low energy particles and in the Klein–Nishina regime for very energetic
    electrons. In particular, the former takes place if the following condition is satisfied $\gamma \epsilon_{ph}/(mc^2) \ll 1$  and in the latter case $\gamma \epsilon_{ph}/(mc^2) \gg 1$, where $\epsilon_{ph} \approx kT$  is the photon’s energy and $T$ denotes the temperature of the accretion disk. Radiation power for the Thomson regime is given by:
    \begin{equation}
        P_T = \frac{\sigma_T\sigma T^4}{4} \frac{\gamma^2}{1+\frac{\gamma k_{B}T}{m_{e}c^2}}
    \end{equation}
    and the radiation power in the  Klein–Nishina regime writes as \citep{Blumenthal1970}:
    \begin{equation}
        P_{KN} \simeq \frac{\sigma_{T}(m_eck_{B}T)^2}{16\hbar^3}\left(\ln{\frac{4\gamma k_BT}{m_ec^2}}-1.981\right)\left(\frac{R_s}{r}\right)^2,
    \end{equation}
where $\sigma_{T}$ is the Thomson cross section, $\sigma$ is the Stefan-Boltzmann constant, $k_{B}$ is Boltzmann constant, $T\approx8.82\times10^{4}$ is the characteristic temperature of the accretion disk at $r=2R_{s}$ \citep{Carroll2017}
\begin{equation}
        T=\left( \frac{GM_{BH}\dot{M}}{8\pi\sigma R_{s}^3} \right)^{\frac{1}{4}}\left(\frac{R_{s}}{r}\right)^{\frac{3}{4}}\left(1-\sqrt{\frac{R_{s}}{r}}\right)^{\frac{1}{4}}
    \end{equation}
where 
\begin{equation}
        \dot{M} = \pi\left(\frac{GM_{BH}}{u_\infty^2}\right)^2\rho_{\infty}u_{\infty},
    \end{equation}
is the accretion rate for the Bondi model \citep{Shapiro1983}, where $R_s$ is the Schwarzschild radius of the black hole. 

    \subsection{Curvature Radiation}
    Even if the field lines are almost straight, in the laboratory frame of reference, their trajectories might be significantly curved, leading to the mechanism of curvature radiation. The energy loss rate by means of the curvature radiation of the electron is given by \citep{Ruderman1975}
    \begin{equation}
        P_{CR} = \frac{2}{3} \frac{e^2 c}{r^2}\gamma^4
    \end{equation}
    The derivation of the maximum Lorentz factor considering the IC effect and curvature radiation is not as trivial as it was in the case of co-rotation constraint. While the acceleration power of the electron is more than the electron emission power, the Lorentz factor of the electron is increasing. Therefore, the Lorentz factor will reach its maximum value when acceleration and emission powers balance each other. The corresponding acceleration power is given by:
    \begin{equation}
        P_{acc} = m_ec^2\frac{d\gamma}{dt}=
        2m_ec\gamma\Omega^2r
        \frac{(1-\frac{\Omega^2r^2}{c^2}-\frac{1}{\gamma^2})^{\frac{1}{2}}}{(1-\frac{\Omega^2r^2}{c^2})}
    \end{equation}
    
\section{Results}
    We describe an electron by the distance between itself and the LC instead of coordinates and by $\gamma$ instead of velocity. From the previous section, we see that all the constraints discussed restrict the position of the electron and the Lorentz factor. In this section, we discuss the possible maximum Lorentz factor for an electron accelerated by different types of black hole.

    To obtain and clarify the results, we constructed graphs (Figs. \ref{M_10}, \ref{fig:M_10000}, \ref{fig:M_1000000} \& \ref{fig:UMBH}) where the so-called allowed and restricted regions are depicted. There are two types of restricted regions: those caused by energy radiation and those caused by the gravitational and magnetic fields of the black hole. We call a region restricted and caused by radiation if $P_{acc} < P_{rad}$, indicating that the electron is losing energy. Conversely, we call a region allowed if $P_{acc} > P_{rad}$, implying that the electron gains energy.

    Restricted regions due to the IC Thomson scattering are coloured blue, while those due to curvature radiation are coloured red. Restrictions caused by co-rotation constraints are depicted in black and differ from those caused by radiation because, in those areas, electrons cannot be confined to magnetic field lines and, therefore, cannot be accelerated by means of the centrifugal acceleration at all. If an electron’s energy exceeds a certain threshold or falls below a certain value, it cannot be confined to the magnetic field line, making acceleration with this model impossible. To neglect the effects of the gravitational field, we explored the acceleration process in the region where $r \gtrapprox 0.6R_{LC}$.
    
    In the next section, we discuss specific values of $\gamma_{max}$ for the respective categories of black holes.
    
    \subsection{Stellar-Mass black holes (Stellar-MBH)}
\begin{figure}
     \centering
     
         \centering
         \includegraphics[width=0.5\textwidth]{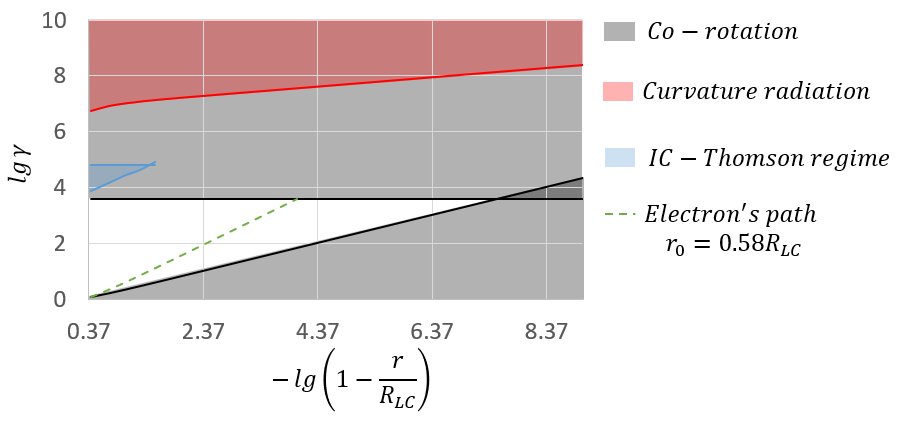}

     \hfill
        \caption{ Map of restricted and allowed regions. Shaded regions represent restrictions with their respective origins. The white region represents allowed Lorentz factors. $M_{BH}=10 M_{\odot}$}
        \label{M_10}
     \end{figure}

    As we have already mentioned, stellar-mass black holes have masses in the range of $1 - 100 M_{\odot}$. The corotation constraint appears to be the only constraining factor that affects $\gamma$ in this type of black hole. In Fig. \ref{M_10}, a black hole with masses of $10 M_{\odot}$ is presented. From the graph, we can see that, in total, all of the constraints can be observed, but only the co-rotation constraint impacts the final Lorentz factor. Thus, regardless of where the electron starts to accelerate, the final $\gamma$ will be the same, approximately $4.1 \times 10^3$ in this case.

\begin{figure}
    \centering
    \includegraphics[width=0.5\textwidth]{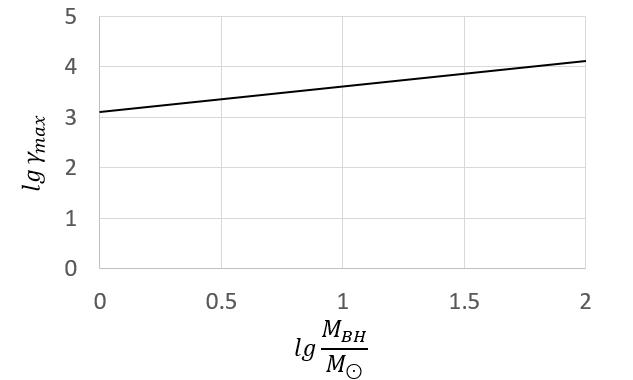}
    \caption{Possible maximum Lorentz factors for Stellar-MBHs}
    \label{fig:Stellar_MBH}
    \end{figure}

    The possible maximum Lorentz factors for black holes of stellar mass are shown in Fig. \ref{fig:Stellar_MBH}, with values ranging from $1.3 \times 10^3$ to $1.3 \times 10^4$.
     
    \subsection{Intermediate-mass black holes (IMBH)}
\begin{figure}
        \centering
        \includegraphics[width=\linewidth]{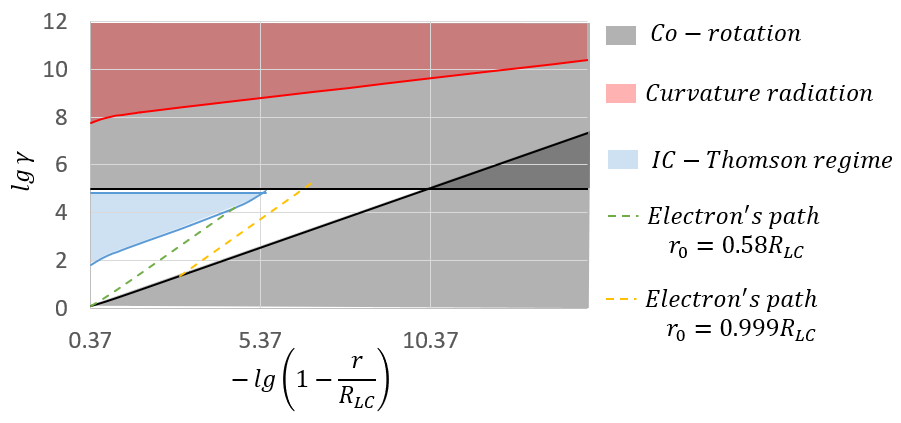}
                \caption{ Map of restricted and allowed regions for $M_{BH}=10^4M_{\odot}$. Shaded regions represent restrictions with their respective origins. White region represents allowed Lorentz factors. } 
        \label{fig:M_10000}
    \end{figure}

    IMBHs are black holes with masses in the range of $10^2 - 10^5$ solar masses.
    If $M_{BH} \lessapprox 10^4 M_{\odot}$, the results are qualitatively similar to those for stellar-MBHs.   
    However, the threshold in this mass range occurs at $M_{BH} \approx 10^4 M_{\odot}$. The IC Thomson regime becomes significant if $M_{BH}$ exceeds this threshold. For example, in Fig. \ref{fig:M_10000}, the results for $M_{BH} = 10^{4} M_{\odot}$ show that all constraining factors impact the outcome, although which factor dominates depends on the initial distance from the black hole. In the case of Fig. \ref{fig:M_10000}, if $0.58 < r_0 < 0.87$, then $\gamma_{max} \approx 1.5\times10^4 -6.7\times10^4$, constrained by the IC Thomson regime. When $r_0 > 0.87$, the IC Thomson effect no longer constrains the electron’s acceleration, allowing it to reach higher Lorentz factors. At these relativistic factors, the electron is constrained by the co-rotation limit, with $\gamma_{max} \approx  10^{5}$. 
    
\begin{figure}
    \centering
    \includegraphics[width=0.5\textwidth]{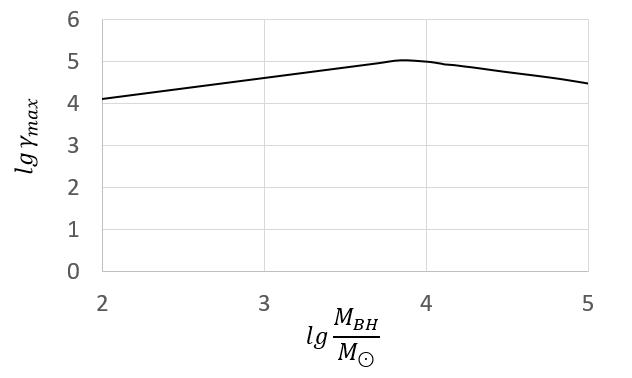} 
    \caption{Possible maximum Lorentz factors for IMBHs}
    \label{fig:IMBH}
    \end{figure}

    It is important to notice on Fig. \ref{fig:IMBH} that after $M_{BH}$ reaches the threshold value of approximately $10^{4}M_{\odot}$, the maximum Lorentz factor begins to decrease. The reason lies behind the expression for the co-rotation constraint, since it is the main constraining factor. From Eq. \ref{Eq:RFmax} we infer that with an increase in mass, two parameters change: the rotational velocity $\Omega$ and the initial relativistic factor $\gamma_{0}$.  In particular, from Eq. \ref{RotVel} $\Omega$ decreases with $M_{BH}$ and $\gamma_{0}$ increases since $r_{0}$ should increase to avoid IC-Thomson Regime and let an electron reach maximal relativistic factor. It appears that after $M_ {BH} \gtrapprox 10^{4}M_{\odot}$ ,  $\gamma_{0}$ increases faster than $\Omega$ decreases therefore the maximal Lorentz factor decreases.

    The complete results for IMBHs are shown in Fig. \ref{fig:IMBH}, from which we can infer that $\gamma_{max} \approx 1.3 \times 10^4 - 1.1 \times 10^{5}$.

    \subsection{Supermassive black holes}

In this section, we present results for the largest group of black holes, with masses in the range of $10^5 - 10^9$ solar masses. For this group, only two constraints affect $\gamma_{max}$: co-rotation and the IC-Thomson Regime. The results for SMBHs with masses $\lessapprox 9\times10^{5}$ are qualitatively the same as for IMBHs with masses over $10^4M_{\odot}$

The threshold in this mass range occurs at $M_{BH} \approx 10^6M_{\odot}$. At this point, the radiation power in the IC Thomson regime becomes so strong that an electron in the Thomson regime cannot accelerate at all. Thus, the only way for an electron to accelerate is if it starts in the Klein-Nishina (KN) regime. In this case, $\gamma \approx 7 \times10^{4}$, which implies that an electron must begin accelerating at almost $r\approx R_{LC}$.

\begin{figure}
    \centering
    \includegraphics[width=0.5\textwidth]{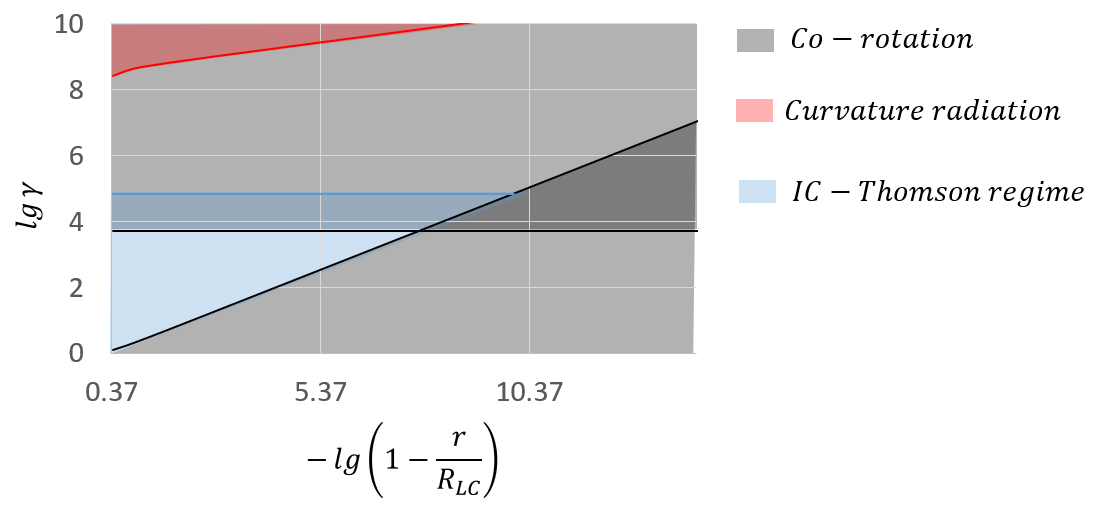} 
    \caption{ Map of restricted and allowed regions. Shaded regions represent restrictions with their respective origins. The white region represents allowed Lorentz factors. $M_{BH} = 10^{6}$ }
    \label{fig:M_1000000}
    \end{figure}

\begin{figure}
        \centering
        \includegraphics[width=\linewidth]{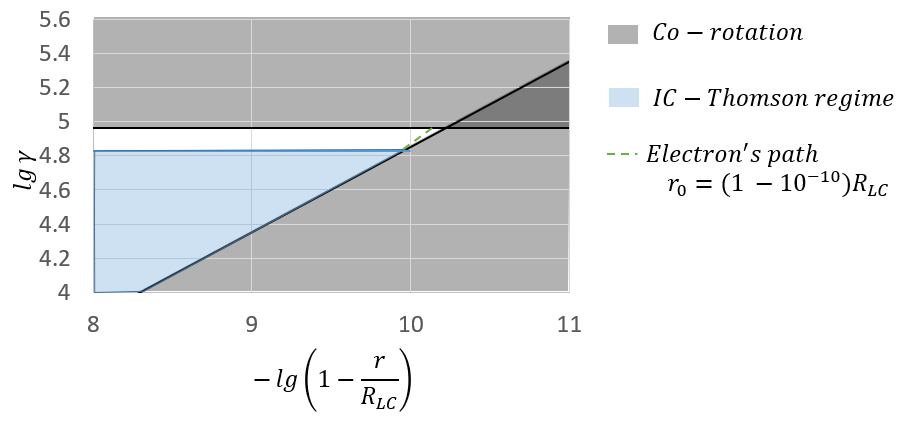}
                \caption{ Map of restricted and allowed regions for $M_{BH}=10^8M_{\odot}$. Shaded regions represent restrictions with their respective origins. The white region represents allowed Lorentz factors. } 
        \label{Fig: M_100000000}
    \end{figure}

 At the same time, if $\gamma_{0} \approx 7 \times 10^{4}$, the maximal relativistic factor set by the co-rotation constraint becomes less than the initial Lorentz factor. In particular, for $M_{BH}=10^{6}M_{\odot}$, $\gamma_{co-rot}=5.3 \times 10^{3} < \gamma_{0}=7\times 10^{4}$ therefore making the acceleration process impossible. This effect is depicted in Fig. \ref{fig:M_1000000}, you can see that there are no white regions and the whole plane is constrained by different constraints; therefore, no acceleration occurs. On the other hand, $\gamma_{0}$ does not increase after increasing in $M_{BH}$ since it is already in the KN regime, but $\gamma_{co-rot}$ does and after $M_{BH}$ surpasses the threshold value of $10^{8}M_{\odot}$, $\gamma_{max}$ becomes larger than the initial Lorentz factor. In particular, when $M_{BH}=3\times10^{8}M_{\odot}$, the maximal relativistic factor is $9.3\times10^{4}$, which is higher than the initial Lorentz factor; therefore, the acceleration process occurs. The effect is depicted in Fig. \ref{Fig: M_100000000}. After $M_{BH}\approx10^{8}M_{\odot}$ maximal Lorentz factor increases since from Eqs. \ref{RotVel} and \ref{Eq:RFmax} it is obvious that $\gamma_{max} \propto M_{BH}^{\frac{1}{2}}$.

\begin{figure}
    \centering
    \includegraphics[width=0.5\textwidth]{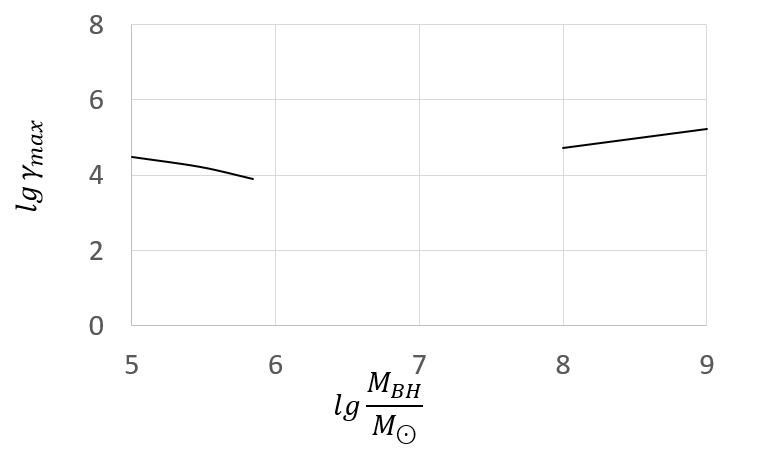} 
    \caption{Possible maximum Lorentz factors for SMBHs}
    \label{fig:SMBH}
    \end{figure}

In Fig. \ref{fig:SMBH}, we can see that the possible maximum Lorentz factors for SMBHs range from $7 \times 10^{4}$ to $1.7 \times 10^{5}$.

    \subsection{Ultramassive black holes}
\begin{figure}
    \centering
    \includegraphics[width=0.5\textwidth]{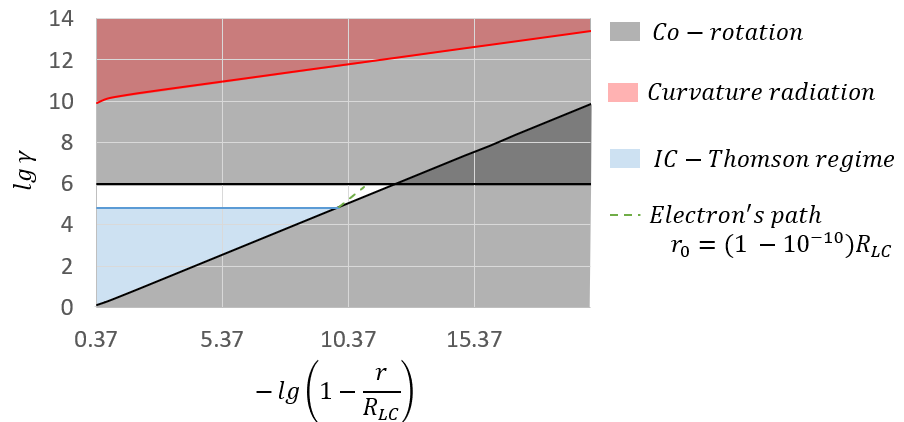} 
    \caption{Map of restricted and allowed zones for electrons in the magnetosphere of the ultramassive black hole in the centre of Abell 1201}
    \label{fig:UMBH}
    \end{figure}

    As we have already mentioned, another category of black holes is sometimes considered for those with masses over $10^9 M_{\odot}$, such as the one at the centre of Abell 1201 with mass $3.3 \times 10^{10} M_{\odot}$. In this section, we present results specifically for that black hole.
    The mass of these black holes is immense, resulting in a low angular velocity. Consequently, the acceleration power is much smaller than the radiation power in the inverse Compton Thomson regime. Thus, the only feasible way for an electron to accelerate is to do so very close to the LC, the same as for SMBHs with masses exceeding $9 \times 10^5 M_{\odot}$, as described in the respective subsection.
    From Fig. \ref{fig:UMBH}, it can be inferred that the maximum relativistic factor is of the order of $10^{6}$

    It should be noted that Lorentz factors for electrons achieve values over $\approx10^6$ and can achieve even higher values for BHs with higher mass. These electrons can produce high-energy radiation, which is the primary focus of our study, therefore making the suggested model of electron acceleration effective for producing high-energy emission of UMBHs.
    
    \subsection{Spin factor}
\begin{figure}
        \centering
        \includegraphics[width=\linewidth]{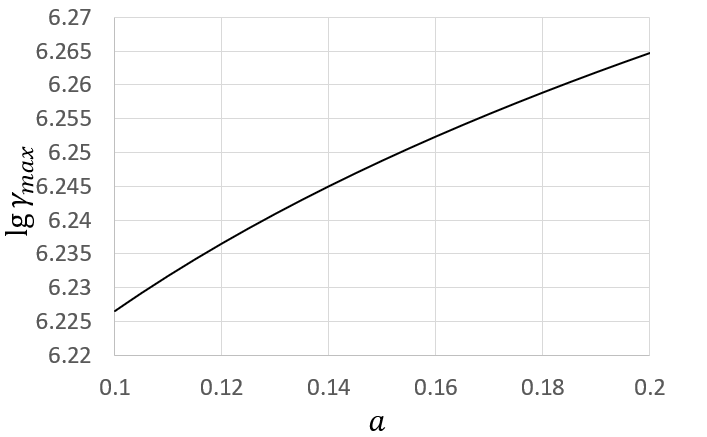}
        \caption{The dependence of maximum relativistic factor on the spin factor for BH with $M_{BH} = 10^{11} M_{\odot}$}
        \label{fig:Spin_factor}
    \end{figure}

    In this section, we explore how the spin factor can influence the maximum Lorentz factor of electrons. We examine relatively small spin parameters to neglect gravitational effects on the electron acceleration process; therefore, we consider the range 0.1 - 0.2. The results for the black hole with mass $M_{BH} = 10^{11}M_{\odot}$ can be seen in Fig. \ref{fig:Spin_factor}. 

    From Fig. \ref{fig:Spin_factor}, it is evident that the relativistic factor is a continuously increasing function of the spin parameter. However, since we examine a relatively narrow range of $a$, the Lorenz factors are not very sensitive to the corresponding change. The increasing character of the relativistic factor is a direct consequence of  Eqs. (\ref{Eq: MagneticField}, \ref{Eq:RFmax}). Since the main constraining factor is co-rotation, at first glance, from Eq. \ref{Eq:RFmax} we can deduce that $\gamma_{max}\sim\Omega^{-\frac{1}{2}}$, but we must not forget that magnetic field induction also depends on the spin factor. From Eq. \ref{Eq: MagneticField}, we can conclude that $B\sim\Omega^{\frac{5}{4}}$. If we take into account the dependence of $B$ on the BH angular velocity, it appears that $\gamma_{max}\sim\Omega^{\frac{1}{8}}$. From both the qualitative analysis and Fig. \ref{fig:Spin_factor}, we conclude that although the change in spin factor affects the maximum Lorentz factor, the increase is very small. If we increase $a$ from 0.1 to 0.2, the logarithm of $\gamma_{max}$ increases by $\lessapprox$  1\% in all of the discussed cases from Fig. \ref{fig:Spin_factor}.
\section{Conclusions}
    The present work examines the centrifugal acceleration as an efficient mechanism for electrons to achieve high energies. We consider electrons confined to a straight magnetic field line, with their acceleration limited by several constraining factors: the inverse Compton effect, curvature radiation, and the breakdown of particle-to-field-line confinement. Our study focuses on investigating how the maximum energy and the main constraining factor depend on the mass of a black hole, with a particular emphasis on IMBHs and UMBHs, as these have not been extensively explored before. Additionally, we study how maximum relativistic factors change due to the initial location of the particle.
\par
    It appears that, even for a given black hole, the maximum energy of electrons depends on where the particle starts the acceleration process (whether closer to or further from the black hole). For stellar-MBHs, the only constraining factor is the co-rotation constraint, and regardless of where the particle begins acceleration, the maximum energy remains the same. For stellar-MBHs, the Lorentz factor is within the range of $1.3 \times 10^3-1.3 \times 10^4-1.3 \times 10^4$.
    
\par
    The case of IMBHs is particularly intriguing as it presents a critical threshold point at $10^{4} M_{\odot}$.  If $M_{BH}$ exceeds $10^{4} M_{\odot}$, particles accelerated closer to the black hole become strongly affected by the inverse Compton scattering in the Thomson regime, requiring particles to start accelerating closer to the LC to achieve maximal energies bound by co-rotation since ones that start closer to BH are bound by IC -in the Thomson regime. On the other hand, an increase in $r_0$ causes an increase in the initial relativistic factor, which in turn decreases the maximal Lorentz factor according to Eq. \ref{Eq:RFmax}. An increase in the initial Lorentz factor for electrons makes Fig. \ref{fig:IMBH} decrease after the threshold mass of $10^{4}M_{BH}$. The range of maximum Lorentz factors is approximately $1.3 \times 10^4-1.1 \times 10^{5}$.
\par
    SMBHs with masses less than $9\times10^5 M_{\odot}$ accelerate particles qualitatively similarly to IMBHs. When the threshold value of  $9\times10^5M_{\odot}$ is reached, the IC Thomson regime becomes so strong that electrons need to have initial Lorentz factors of the order of $10^4$, therefore need to be confined very close to the LC, to start the acceleration process. Unfortunately, a high initial Lorentz factor causes the co-rotation constraint to be very low, therefore making acceleration impossible until the mass of the BH is so large that the co-rotation constraint becomes large again, making acceleration achievable again. This leads us to the second threshold mass in the group of BHs at $10^{8}$ where a large mass causes the rotational velocity of BH to be so small that it counteracts to large initial Lorentz factor and makes centrifugal force acceleration an efficient mechanism for electrons to gain energies. For SMBHs, the range of maximum Lorentz factors spans from $7 \times 10^{4}$ to $1.7 \times 10^{5}$. 
\par
    As an example of a UMBH, we explore the ultra-massive black hole at the centre of Abell 1201, where the relativistic factors are of the order of $ 10^{6}$. It is important to notice that in this group of black holes, the mechanism becomes efficient in producing high-energy electrons.
\section*{Acknowledgements}

The research was supported by the Shota Rustaveli National Science Foundation of Georgia (SRNSFG). Grant: FR-24-1751. The work of N.N. was supported by the Knowledge Foundation of the Free University of Tbilisi.

\section*{Data Availability}

Data are available in the article and can be accessed via a DOI link.



\bibliographystyle{mnras}
\bibliography{mnras_template} 





\bsp	
\label{lastpage}
\end{document}